\documentclass[cits]{PoS}
\usepackage[mathscr]{euscript}

\def\LP{\left(}		
\def\RP{\right)}	

\def\BNE{\begin{equation}}
\def\ENE{\end{equation}}
\def\BE{\begin{displaymath}}
\def\EE{\end{displaymath}}
\def\BEA{\begin{eqnarray*}}
\def\EEA{\end{eqnarray*}}
\def\BNEA{\begin{eqnarray}}
\def\ENEA{\end{eqnarray}}

\newcommand{\ie}{{\em i.e.\ }}
\newcommand{\eg}{{\em e.\,g.\ }}

\newcommand{\etc}{etc.\ }

\definecolor{Red}           {cmyk}{0,1,1,0}
\definecolor{Blue}          {cmyk}{1,1,0,0}
\definecolor{Green}         {cmyk}{1,0,1,0}

\newcommand{\R}[1]{{\color{Red}#1}}

\newcommand{\B}[1]{{\color{Blue}#1}}

\def\chpt{\raise0.4ex\hbox{$\chi$}PT}
\def\schpt{S\raise0.4ex\hbox{$\chi$}PT}
\def\rschpt{rS\raise0.4ex\hbox{$\chi$}PT}
\def\figref#1{Fig.~\ref{fig:#1}}
\def\Figref#1{Figure~\ref{fig:#1}}

\def\tabref#1{Table~\ref{tab:#1}}

\def\gtwid{{\,\raise.3ex\hbox{$>$\kern-.75em\lower1ex\hbox{$\sim$}}\,}}
\def\ltwid{{\,\raise.3ex\hbox{$<$\kern-.75em\lower1ex\hbox{$\sim$}}\,}}

\def\ie{{\it i.e.},\ }
\def\eg{{\it e.g.},\ }
\def\et{{\it et al.}}
\def\etc{{\it etc.}\ }

\def\cM{{\cal M}}

\def\cO{{\cal O}}

\def\rcite#1{Ref.~\cite{#1}}

\def\rcites#1{Refs.~\cite{#1}}

\def\eqn#1{\label{eq:#1}}

\def\eq#1{Eq.~(\ref{eq:#1})}
\def\eqsthru#1#2{Eqs.~(\ref{eq:#1}) through (\ref{eq:#2})}

\def\aschpt{HMrAS\raise0.4ex\hbox{$\chi$}PT}

\title{Charmed and strange pseudoscalar meson decay constants from HISQ simulations}

\ShortTitle{Charmed and strange pseudoscalar meson decay constants from HISQ simulations}

\author{
A.~Bazavov$^a$
  \hspace{-1.0mm}\thanks{Present address: Department of Physics and Astronomy, University of Iowa, Iowa City, IA
52245},
\speaker{C.~Bernard} \nolinebreak $^b$,
C.~Bouchard$^c$,
C.~DeTar$^d$,
D.~Du$^e$,
A.X.~El-Khadra$^{e,f}$,
J.~Foley$^d$,
E.D.~Freeland$^g$,
E.~G\'amiz$^h$,
Steven~Gottlieb$^i$,
U.M.~Heller$^j$,
J.~Kim$^k$
    \hspace{-1.0mm}\thanks{Present address: Lattice Gauge Theory Research Center, CTP, and FPRD, Department
of Physics and Astronomy, Seoul National University, Seoul, 151-747, South Korea},
J.~Komijani$^b$,
A.S.~Kronfeld$^f$,
J.~Laiho$^l$
    \hspace{-1.0mm}\thanks{Present address: Department of Physics, Syracuse University, Syracuse, New York,
USA},
L.~Levkova$^d$,
P.B.~Mackenzie$^f$,
E.T.~Neil$^f$
    \hspace{-1mm}\thanks{Present address: Department of Physics, University of Colorado, Boulder, CO 80309, USA
and RIKEN-BNL Research Center, Brookhaven National Laboratory, Upton, NY 11973, USA} ,
J.N.~Simone$^f$,
R.L.~Sugar$^m$,
\speaker{D.~Toussaint}\nolinebreak $^k$,
R.S.~Van~de~Water$^f$,
and
R.~Zhou$^i$
 \hspace{-1mm}\thanks{Present address: Fermi National Accelerator Laboratory, Batavia, IL 60510, USA}
\\
\llap{$^a$} Department of Physics, Brookhaven National Laboratory\thanks{Operated by Brookhaven Science Associates, LLC, under
Contract No.~DE-AC02-98CH10886 with
the U.S. Department of Energy.}, \ $\;$Upton, NY 11973, USA\\
\llap{$^b$} Department of Physics, Washington University, St. Louis, MO 63130, USA\\
\llap{$^c$} Department of Physics, The Ohio State University, Columbus, OH 43210, USA\\
\llap{$^d$}  Department of Physics and Astronomy, University of Utah, Salt Lake City, UT 84112, USA\\
\llap{$^e$} Physics Department, University of Illinois, Urbana,  IL 61801, USA\\
\llap{$^f$} Fermi National Accelerator Laboratory\thanks{Operated by Fermi Research Alliance, LLC, under Contract
No.~DE-AC02-07CH11359 with
the U.S. Department of Energy.}, \ $\;$Batavia, IL 60510, USA\\
\llap{$^g$} Department of Physics, Benedictine University, Lisle, IL 60532, USA\\
\llap{$^h$} CAFPE and Departamento de F\'isica Te\'orica y del Cosmos, Universidad de Granada, Granada, Spain\\
\llap{$^i$} Department of Physics, Indiana University, Bloomington, IN 47405, USA\\
\llap{$^j$} American Physical Society, One Research Road, Ridge, NY 11961, USA\\
\llap{$^k$} Physics Department, University of Arizona, Tucson, AZ 85721, USA\\
\llap{$^l$} SUPA, School of Physics and Astronomy, University of Glasgow, Glasgow G12 8QQ, UK\\
\llap{$^m$} Department of Physics, University of California, Santa Barbara, CA 93106, USA

\vspace{2mm}
{\large\bf Fermilab Lattice and MILC Collaborations}
\vspace{3mm}

E-mail:
\email{doug@physics.arizona.edu,cb@wustl.edu}
}

\abstract{
We update our determinations of $f_{D^+}$, $f_{D_s}$, $f_K$, and quark mass ratios
from simulations with four flavors of HISQ dynamical quarks.
The availability of ensembles with light quarks near their physical
mass means that we can extract physical results with only
small corrections for valence- and sea-quark mass mistunings
instead of a chiral extrapolation.
The adjusted valence-quark masses and lattice spacings  may be determined from
an ensemble-by-ensemble analysis, and the results for the quark mass ratios
then extrapolated to the continuum limit. Our central values
of the charmed meson decay constants, however, come from an alternative analysis,
which uses staggered chiral perturbation theory for the heavy-light mesons, and
allows us to incorporate data at unphysical quark
masses where statistical errors are often smaller.  
A jackknife analysis propagated through all of these steps takes account of the
correlations among all the quantities used in the analysis.  Systematic errors
from the finite spatial size and EM effects are estimated by varying the parameters
in the analysis, and systematic errors from the assumptions in the continuum extrapolation
are estimated from the spread of values from different extrapolations.
}

\FullConference{31st International Symposium on Lattice Field Theory - LATTICE 2013\\
		July 29 - August 3, 2013\\
		Mainz, Germany}

\begin{document}

\section{Introduction}

The pseudoscalar meson decay constants $f_K$, $f_{D^+}$ and $f_{D_s}$, together with experimental
decay rate determinations, are the simplest, although not necessarily most
precise, ways to determine $V_{us}$, $V_{cd}$ and $V_{cs}$.
These decay constants are obtained from the amplitude of a point-source and
point-sink (or random wall equivalent) pseudoscalar correlator $A_{pt\mbox{-}pt}$
\BNE\label{eq:fpi_formula} f_{pseudo} = \LP m_A + m_B \RP \sqrt{ \frac{3 V A_{pt\mbox{-}pt}}{2 M_{pseudo}^3} } \ \ \ ,\ENE
where $m_A$ and $m_B$ are the quark masses, and $V$ is the spatial volume.

Here we update our determinations of these decay constants, and the quark mass ratios
that are also produced in the analysis, from our program of simulations using a one-loop
Symanzik improved gauge action and the highly-improved staggered quark (HISQ) action \cite{HPQCD_HISQ}.
The HISQ action reduces taste violation errors by roughly a factor of three compared
with the asqtad action.  Also, the charm quark dispersion relation is improved so
that charm quarks can be treated with the same relativistic action as the light quarks.
Our ensembles include four flavors of dynamical quarks.  Although the effects of
a sea charm quark are expected to be small, the cost of including them is very small.

\section{Correlator masses and amplitudes}\vspace{-2.0mm} \label{CORRELATORS}

The first step in our analysis is to find the masses and amplitudes of the
pseudoscalar meson correlators, $M_{pseudo}$ and  $A_{pt\mbox{-}pt}$ in Eq.~(\ref{eq:fpi_formula}).
Our ensembles of lattices and many of their characteristics were presented
in Ref.~\cite{HISQ_CONFIGS}.  Table~\ref{tab:ensembles} shows the ensembles and
numbers of lattices used in this analysis, while Table~\ref{tab:valencemasses}
shows the valence masses used in each ensemble.
Our sources, and choices of fitting functions and fit ranges were presented in Ref.~\cite{KIM_LATTICE12},
so we simply present them without further discussion in Table~\ref{tab:fittypes}.

\begin{table}
\caption{ \label{tab:ensembles}
Primary lattice ensembles used in this calculation.  These ensembles have $m_s$ tuned close to its physical value.  A $*$ in the $N_{lats}$ column indicates that  lattice generation is still in progress. $m_\pi$ is in MeV. Ensembles with $m_s$ less than its physical value are also included
in the chiral fit in Sec.~4.   
}
\begin{center}\begin{tabular}{|l|lll|l|l|llll|}
\hline
$\beta$ & $am_l$ & $am_s$ & $am_c$ & size      & $N_{lats}$ & $a$ (fm)& $L$ (fm) & $m_\pi L$ & $m_\pi$\\ \hline
\hline
5.80  & 0.013   & 0.065   & 0.838 & $16^3\times 48$  & 1020  & 0.14985(38) & 2.38 & 3.8 & 314 \\
5.80  & 0.0064  & 0.064   & 0.828 & $24^3\times 48$  & 1000  & 0.15303(19) & 3.67 & 4.0 & 214 \\
5.80  & 0.00235 & 0.0647  & 0.831 & $32^3\times 48$  & 1000  & 0.15089(17) & 4.83 & 3.2 & 130 \\
\hline
6.00  & 0.0102  & 0.0509  & 0.635 & $24^3\times 64$  & 1040  & 0.12520(22) & 3.00 & 4.5 & 299 \\
6.00  & 0.00507 & 0.0507  & 0.628 & $24^3\times 64$  & 1020  & 0.12085(28) & 2.89 & 3.2 & 221 \\
6.00  & 0.00507 & 0.0507  & 0.628 & $32^3\times 64$  & 1000  & 0.12307(16) & 3.93 & 4.3 & 216 \\
6.00  & 0.00507 & 0.0507  & 0.628 & $40^3\times 64$  & 1028  & 0.12388(10) & 4.95 & 5.4 & 214 \\
6.00  & 0.00184 & 0.0507  & 0.628 & $48^3\times 64$  & 999   & 0.12121(10) & 5.82 & 3.9 & 133 \\
\hline
6.30  & 0.0074  & 0.037   & 0.440 & $32^3\times 96$  & 1011  & 0.09242(21) & 2.95 & 4.5 & 301 \\
6.30  & 0.00363 & 0.0363  & 0.430 & $48^3\times 96$  & 1000  & 0.09030(13) & 4.33 & 4.7 & 215 \\
6.30  & 0.0012  & 0.0363  & 0.432 & $64^3\times 96$  & 1031  & 0.08773(08) & 5.62 & 3.7 & 130 \\
\hline
6.72  & 0.0048  & 0.024   & 0.286 & $48^3\times 144$ & 1016  & 0.06132(22) & 2.94 & 4.5 & 304 \\
6.72  & 0.0024  & 0.024   & 0.286 & $64^3\times 144$ & 1166  & 0.05938(12) & 3.79 & 4.3 & 224 \\
6.72  & 0.0008  & 0.022   & 0.260 & $96^3\times 192$ & 583*  & 0.05678(06) & 5.44 & 3.7 & 135 \\
\hline
\end{tabular}\end{center}
\end{table}

\begin{table}
\caption{ \label{tab:valencemasses}
Valence masses used on each ensemble.
The sea-quark masses $am_l$, $am_s$ and $am_c$ are in lattice units, while the
valence-quark masses are given as fractions of the sea strange or charm quark
mass.
}
\begin{center}\begin{tabular}{|l|lll|r|r|}
\hline
$\beta$ & $am_l$ & $am_s$ & $am_c$ & light masses $m_A$ & $m_B$     \\
        &        &        &        &          ($m/m_s$) & ($m/m_c$) \\
\hline
5.80  & 0.013   & 0.065   & 0.838 & 0.1,0.15,0.2,0.3,0.4,0.6,0.8,1.0 & 0.9,1.0 \\
5.80  & 0.0064  & 0.064   & 0.828 & 0.1,0.15,0.2,0.3,0.4,0.6,0.8,1.0 & 0.9,1.0 \\
5.80  & 0.00235 & 0.0647  & 0.831 & 0.036,0.07,0.1,0.15,0.2,0.3,0.4,0.6,0.8,1.0& 0.9,1.0 \\
\hline
6.00  & 0.0102  & 0.0509  & 0.635 & 0.1,0.15,0.2,0.3,0.4,0.6,0.8,1.0 & 0.9,1.0 \\
6.00  & 0.00507 & 0.0507  & 0.628 & 0.1,0.15,0.2,0.3,0.4,0.6,0.8,1.0 & 0.9,1.0 \\
6.00  & 0.00507 & 0.0507  & 0.628 & 0.1,0.15,0.2,0.3,0.4,0.6,0.8,1.0 & 0.9,1.0 \\
6.00  & 0.00507 & 0.0507  & 0.628 & 0.1,0.15,0.2,0.3,0.4,0.6,0.8,1.0 & 0.9,1.0 \\
6.00  & 0.00507 & 0.0304  & 0.628 & 0.1,0.15,0.2,0.3,0.4,0.6,0.8,1.0 & 0.9,1.0 \\
6.00  & 0.00507 & 0.00507 & 0.628 & 0.1,0.15,0.2,0.3,0.4,0.6,0.8,1.0 & 0.9,1.0 \\
6.00  & 0.00184 & 0.0507  & 0.628 & 0.036,0.073,0.1,0.15,0.2,0.3,0.4,0.6,0.8,1.0 & 0.9,1.0 \\
\hline
6.30  & 0.0074  & 0.037   & 0.440 & 0.1,0.15,0.2,0.3,0.4,0.6,0.8,1.0 & 0.9,1.0 \\
6.30  & 0.00363 & 0.0363  & 0.430 & 0.1,0.15,0.2,0.3,0.4,0.6,0.8,1.0 & 0.9,1.0 \\
6.30  & 0.0012  & 0.0363  & 0.432 & 0.033,0.066,0.1,0.15,0.2,0.3,0.4,0.6,0.8,1.0 & 0.9,1.0 \\
\hline
6.72  & 0.0048  & 0.024   & 0.286 & 0.05,0.1,0.15,0.2,0.3,0.4,0.6,0.8,1.0 & 0.9,1.0 \\
6.72  & 0.0024  & 0.024   & 0.286 & 0.05,0.1,0.15,0.2,0.3,0.4,0.6,0.8,1.0 & 0.9,1.0 \\
6.72  & 0.0008  & 0.022   & 0.260 & 0.036,0.068,0.1,0.15,0.2,0.3,0.4,0.6,0.8,1.0 & 0.9,1.0 \\
\hline
\end{tabular}\end{center}
\end{table}

\begin{table}
\caption{ \label{tab:fittypes}
Fit forms and minimum distance included for the two-point correlator
fits.  Here the fit form is the number of negative parity
(\ie pseudoscalar) states ``plus'' the number of positive parity
states.  In cases when the valence quarks have equal masses, the
opposite parity states are not present.
In this work, the charm-charm fits are used only in computing the
mass of the $\eta_c$ meson, which serves as a check on the quality of our
charm physics.
}
\begin{center}\begin{tabular}{|l|ll|ll|ll|}
\hline
                & \multicolumn{2}{c|}{light-light} &
\multicolumn{2}{c|}{light-charm} &
\multicolumn{2}{c|}{charm-charm} \\
\hline
                & form & $D_{min}$ & form & $D_{min}$ & form & $D_{min}$ \\
\hline
$a\approx 0.15$ fm & 1+1 & 16 & 2+1 & 8 & 2+0 & 9 \\
$a\approx 0.12$ fm & 1+1 & 20 & 2+1 & 10 & 2+0 & 12 \\
$a\approx 0.09$ fm & 1+1 & 30 & 2+1 & 15 & 2+0 & 18 \\
$a\approx 0.06$ fm & 1+1 & 40 & 2+1 & 20 & 2+0 & 21 \\
$a\approx 0.045$ fm & 1+1 & 53 & 2+1 & 26 & 2+0 & 31 \\
\hline
\end{tabular}\end{center}
\end{table}

\section{Lattice spacing, quark masses and decay constants on each ensemble}\vspace{-2.0mm} \label{PERENSEMBLE}
\label{sec:by_ensemble}

\begin{figure}
\vspace{-0.50in}
\begin{center}\includegraphics[width=0.65\textwidth]{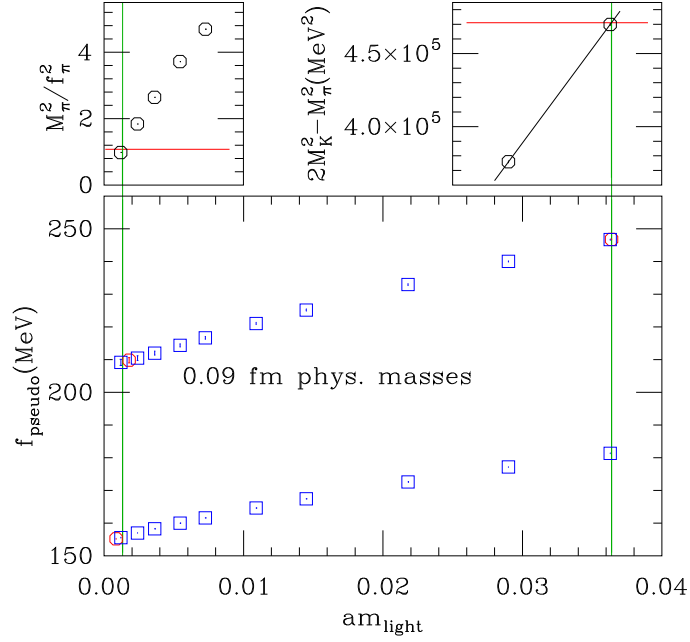}\end{center}
\vspace{-0.95in}
\caption{ \label{fig:tuning}
Illustration of the lattice spacing and quark mass tuning. The text
contains a discussion of the procedure.  This example is for the
$a\approx 0.09$ fm ensemble with light quark mass near its physical value.
}
\vspace{-0.15in}
\end{figure}

After determining the amplitudes and masses for the pseudoscalar correlators for
all of the valence-quark masses, we determine the lattice spacing
and corrected quark masses on each ensemble.
Figure~\ref{fig:tuning} illustrates the first steps in this procedure.
We begin by finding the valence mass where $M_\pi^2/f_\pi^2$ has its
physical value (adjusted for finite size effects), which is illustrated
in the upper left panel of Fig.~\ref{fig:tuning}. Here the red horizontal line
is the desired value, and the green vertical line, the resulting quark mass.
In solving for this point, we use the NLO continuum chiral perturbation
theory form for $M_\pi/f_\pi$ and linear and quadratic analytic terms.
Using $f_\pi = 130.41$ MeV, we fix the lattice spacing $a$.
Then, interpolating in light quark mass to the value found above, we
use ``kaons'' with strange valence mass at 1.0 and 0.8 times the sea
strange quark mass.  The quantity $2M_K^2-M_\pi^2$ is linearly
interpolated or extrapolated to the point where it reaches its physical
value to determine the adjusted strange quark mass, as illustrated in
the upper right panel of Fig.~\ref{fig:tuning}.  
Then (not illustrated in the figure), we perform a similar interpolation
in $M_{D_s}$ to determine the adjusted charm quark mass.
Finally the difference of light quark masses $m_d-m_u$ is determined
from the $K_0$--$K^+$ mass splitting after adjustments for electromagnetic
effects \cite{EM_EFFECTS}.
The use of these masses to determine $f_K$, $f_{D^+}$ and $f_{D_s}$ is illustrated
in the lower panel of Fig.~\ref{fig:tuning}.  This shows the decay constants
as a function of light quark mass, where the other quark mass has been
interpolated or extrapolated to adjust the strange (lower points) or
charm (upper points) mass determined above.  We then interpolate
in these points to light quark mass $m_u$, $m_d$ or $m_s$ to find
$f_K$, $f_{D^+}$ or $f_{D_s}$ respectively.

\section{Chiral perturbation theory analysis of $f_{D^+}$ and $f_{D_s}$}\vspace{-2.0mm} \label{CHIPT}

The quark-mass and lattice-spacing dependence of the decay constant has been calculated at one loop in  heavy-meson, rooted, all-staggered chiral perturbation theory (\aschpt) in \rcite{Bernard-Komijani}.
At fixed heavy-quark mass $m_Q$,  one may argue 
following Ref.~\cite{FermilabMILC_Dec2011} and using
the power counting introduced by Boyd and Grinstein \cite{BoydGrinstein} that inclusion of 
hyperfine splittings
(\eg $M^*_D-M_D$) and flavor splittings (\eg $M_{D_s}-M_D$), but no other $1/m_Q$ effects,
constitutes a systematic approximation at NLO in \aschpt. 
With ${\rm v}$ denoting the light valence quark, $X$ the ${\rm v}\bar {\rm v}$ valence meson, and
$\Phi_{D_{\rm v}}\equiv f_{D_{\rm v}} \sqrt{ M_{D_{\rm v}}}$,
\rcite{Bernard-Komijani} obtains for the pseudoscalar-taste heavy-light meson:
\begin{eqnarray}\label{eq:chpt-form}
 \Phi_{D_{\rm v}}
      & = &     \Phi_0 \Biggl\{  1 + \frac{1}{16\pi^2f^2} \frac{1}{2}
      \Biggl(-\frac{1}{16}\sum_{\mathscr{S},\Xi} \ell(M_{\mathscr{S}{\rm v},\Xi}^2)
          - \frac{1}{3}
          \sum_{j\in \cM_I^{(3,{\rm v})}} 
          \frac{\partial}{\partial M^2_{X,I}}\left[ 
                R^{[3,3]}_{j}(
              \cM_I^{(3,{\rm v})};  \mu^{(3)}_I) \ell(M_{j}^2) \right]
              \nonumber \\*&&\hspace{-5mm}{} 
              -   \Bigl( a^2\delta'_V \sum_{j\in \cM_V^{(4,{\rm v})}}
              \frac{\partial}{\partial M^2_{X,V}}\left[ 
                R^{[4,3]}_{j}( \cM_V^{(4,{\rm v})}; \mu^{(3)}_V)
              \ell(M_{j}^2)\right]
                  + [V\to A]\Bigr) \nonumber \\*&&\hspace{-5mm}{}
            -3g_\pi^2\frac{1}{16}\sum_{\mathscr{S},\Xi} J(M_{\mathscr{S}{\rm v},\Xi},\Delta^*+\delta_{\mathscr{S}{\rm v}})
            - g_\pi^2
          \sum_{j\in \cM_I^{(3,{\rm v})}} 
          \frac{\partial}{\partial M^2_{X,I}}\left[ 
                R^{[3,3]}_{j}(
              \cM_I^{(3,{\rm v})};  \mu^{(3)}_I) J(M_{j},\Delta^*) \right]
              \nonumber \\*&&\hspace{-5mm}{} 
              \hspace{0cm} -3g_\pi^2 \Bigl( a^2\delta'_V \sum_{j\in \cM_V^{(4,{\rm v})}}
              \frac{\partial}{\partial M^2_{X,V}}\left[ 
                R^{[4,3]}_{j}( \cM_V^{(4,{\rm v})}; \mu^{(3)}_V)
              J(M_{j},\Delta^*)\right]
                  + [V\to A]\Bigr)  
            \Biggr)\ \nonumber \\*&&\hspace{-5mm}{}+
      L_s (x_u + x_d + x_s) + L_{\rm v} x_{\rm v} + L_{a} \frac{x_{\bar\Delta}}{2}\Biggr\} \ , \eqn{chiral-form}
\end{eqnarray}
where $\Phi_0$, $L_s$, $L_{\rm v}$, and $L_a$ are low energy constants (LECs); the indices $\mathscr{S}$ and $\Xi$ run over sea-quark flavors and meson tastes, respectively;
$\Delta^*$ is the lowest-order hyperfine splitting;  $\delta_{\mathscr{S}{\rm v}}$  is the
flavor splitting between a heavy-light meson with  light quark of flavor $\mathscr{S}$ and one
of flavor ${\rm v}$; and $g_\pi$ is the $D$-$D^*$-$\pi$ coupling.
 The chiral logarithm functions $\ell$ and $J$, and the residue functions $R^{[n,k]}_j$ and mass sets
 $\cM^{(3,{\rm v})}$, $\cM^{(4,{\rm v})}$, and $\mu^{(3)}$,  are defined in \rcites{Bernard-Komijani,Aubin:StagHL2007}.  Subscripts 
 on these mass sets indicate the taste.
We define dimensionless quark 
masses and a measure of the taste splitting by
\begin{eqnarray}\eqn{x-defs}
x_{u,d,s,{\rm v}}  \equiv \frac{4B}{16 \pi^2 f_{\pi}^2} m_{u,d,s,{\rm v}}\;, &\hspace{15mm}&
x_{\bar\Delta} \equiv \frac{2}{16 \pi^2 f_{\pi}^2}\bar\Delta \;,
\end{eqnarray}
where $B$ is the LEC that gives the Goldstone pion mass $M_\pi^2 = B(m_u+m_d)$, and
$\bar\Delta$ is the mean-squared pion taste splitting.  The $x_i$ are the natural variables of \aschpt; the LECs $L_s$, $L_{\rm v}$, and $L_a$ are therefore expected to be $\cO(1)$. All ensembles in the current analysis have degenerate light quarks: $x_u=x_d\equiv x_l$. The taste splittings have been determined
to $\sim\!1$--10\% precision \cite{HISQ_CONFIGS} and are used as input to \eq{chiral-form}, as are
the taste-breaking hairpin parameters $\delta'_A$ and $\delta'_V$, whose ranges
are taken from chiral fits to light pseudoscalar mesons \cite{Bazavov:2011fh}. 

While \eq{chiral-form} is a systematic NLO approximation for the decay constant at fixed $m_Q$, we have data on each ensemble with two different values of the (valence) charm mass: $m_c$ and $0.9 m_c$,
where $m_c$ is the value of the sea charm mass of the ensembles, and is itself not precisely equal to the physical charm mass $m_c^{\rm phys}$ because of tuning errors. In order to fit this data, we allow the LEC $\Phi_0$ to depend on $m_Q$ as suggested by HQET;  for acceptable
fits we need to introduce $1/m_Q$ and $1/m_Q^2$ terms.  Furthermore, $\Phi_0$ has generic lattice-spacing dependence
that must be included to obtain good fits.  With HISQ quarks,
the leading generic discretization errors are $\cO(\alpha_s a^2)$.  But because the high degree of 
improvement in the HISQ action drastically reduces the coefficient of these leading errors, formally higher  $\cO(a^4)$ errors are also apparent. (See, for example, $f_K/f_\pi$ {\it vs.}\/ $a^2$ in \rcite{FKOFPI_MILC13}.)

In \eq{chiral-form}, we thus 
replace 
\begin{equation}\eqn{Phi0-form}
\Phi_0\to \Phi_0 \left(1 + k_1\frac{\Lambda_{\rm QCD}}{m_Q}+ k_2\frac{\Lambda_{\rm QCD}^2}{m_Q^2}\right)\Big(1+ c_1\alpha_s a^2 +c_2  a^4 
\Big)\;,
\end{equation}
where the $k_i$ are new physical LECs, and $c_i$ are additional fit parameters.  In cases where the valence and sea values of the charm quark mass differ, $m_Q$ in
\eq{Phi0-form} is taken to denote the valence mass. Dependence on the charm sea mass over the $\sim\!\!10\%$ range of
variation of $m_Q$ is probably extremely small, since one expects that even the existence of a charm quark in the sea (\ie the difference between 2+1+1 and 2+1 simulations) is a small ($\ltwid\!1\%$) effect.

Generic dependence on $a$ is also allowed 
for the physical LECs $L_s$, $L_{\rm v}$,   $k_1$ and $k_2$.  However, because these parameters first 
appear at  NLO in the chiral or HQET expansions, it is sufficient to include only the leading $a$-dependence, for example:
\begin{equation}\eqn{L-form}
L_{\rm v} \to L_{\rm v} + L_{\rm v\delta} \alpha_s a^2 
\end{equation}
Thus we add 4 fit parameters related to generic discretization effects: $L_{\rm v\delta} $,   $L_{\rm s\delta} $,    $k_{1\delta} $,  and $k_{2\delta}$. 
There are also 3 parameters related to taste-violation effects: $L_a$, $\delta'_A$ and $\delta'_V$.  These parameters are
taken proportional to the measured average taste splitting $\bar \Delta$, which depends on $a$ approximately as
$\alpha_s^2 a^2$  \cite{HISQ_CONFIGS}. In addition, we find that $m_Q$-dependent discretization errors 
must  be considered if data at the coarsest lattice spacing ($a\approx0.15\;$fm) is included in the fits.
This is not surprising because $am^{\rm phys}_c\approx0.85$ at this lattice spacing, which by the power counting
estimates of \rcite{HPQCD_HISQ} suggests $\sim\!5\%$ discretization errors 
(although this may be reduced somewhat by dimensionless factors).  
We therefore add $c_3\alpha_s (am_Q)^2 +c_4(am_Q)^4$ to the analytic terms in \eq{chiral-form}.  If the $a\approx0.15\;$fm
data is omitted, good fits may be obtained with $c_3$ and $c_4$ set to zero.

For the LEC $g_\pi$, a reasonable range is $g_\pi=0.53(8)$, which comes from recent lattice calculations 
 \cite{Becirevic,Detmold}.
However, when this central value and range are included as Bayesian priors, fits to our full data set
 tend to pull  $g_\pi$ low, several sigma below 0.53. Hence, we simply fix 
 $g_\pi=0.45$, 1-sigma below its nominal value, in our current central fits.  This problem is ameliorated for 
 alternative  fits, used to estimate the systematic errors, that drop
 the data at  $a\approx0.15$ fm.   Other alternatives considered are to allow $g_\pi$ to be a free parameter, or to keep it fixed at its nominal value.

Because we have precise data ($\sim\!0.2\%$ statistical errors) and 314 to 366 data points (depending on whether $a\approx0.15\;$fm is included), NLO \aschpt\ is not adequate to describe the quark mass dependence, in  particular for masses near $m_s$.  We therefore include all NNLO and NNNLO mass-dependent analytic terms.  There are 4 independent  functions of $x_{\rm v}$,
$x_l$ and $x_s$ at NNLO and 7 at NNNLO, for a total of 11 additional parameters.  Our central fit then has 28 fit parameters, while alternative fits have between 27 and 31 parameters. 

Relative scale setting is done using $f_{p4s}$ \cite{HISQ_CONFIGS}, which is the light-light 
pseudoscalar decay constant at a fiducial point with both valence masses
equal to $m_{p4s}\equiv0.4m_s^{\rm phys}$, and with the sea-quark masses physical. It is determined in lattice units on the physical-mass ensembles,
with small adjustments for mistunings using nearby ensembles.
 We use a mass-independent scale-setting scheme:  by definition, 
all ensembles
at the same $\beta$ have the same scale as the physical-mass ensemble.
The ratio  $f_{p4s}/M_{p4s}$, where $M_{p4s}$ is the meson mass at the fiducial point, is used to
(re-)tune $m_s$ to its physical value.  Values of the ratio and of $f_{p4s}$ in physical units come from the
analysis on the physical ensembles only, as described above.  That analysis also gives the needed 
quark mass ratios $m_c/m_s$, $m_s/m_l$ and $m_u/m_d$.

\begin{figure}[t]
\null\vspace{-45mm}
\hspace{-8mm}\includegraphics[width=16.2cm]{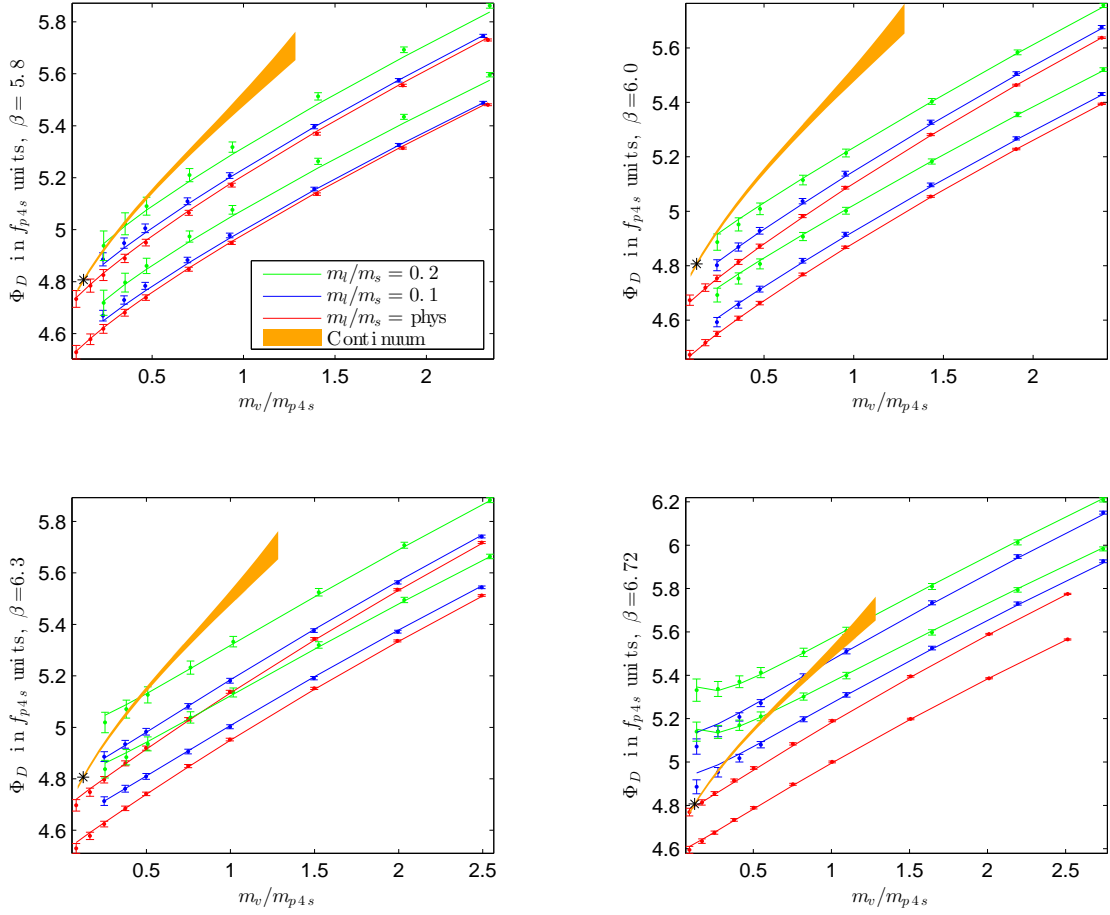}
\vspace{-46mm}
\caption{Simultaneous chiral fit to $\Phi_D$  as a function of $m_{\rm v}$, the valence-quark mass (in
units of $m_{p4s}$), at all four lattice spacings: $a\approx0.15$ fm and $0.12$ fm (top row), and
$0.09$ fm and $0.06$ fm (bottom row).  The colors denote different light sea-quark masses, as indicated.  For each color
there are two lines, one for heavy valence-quark mass $\approx m_c$ (higher line), and one for $\approx 0.9 m_c$. 
This fit has $\chi^2/{\rm dof}=343.5/338$, giving $p=0.41$. The orange
band (which is identical in each plot, although the  vertical and horizontal scales differ) gives the result after extrapolation to the continuum
and adjustment of the strange and charm masses to their physical values.
 The width of the
band  shows the statistical error coming from the fit.
The black bursts indicate the value of $\Phi_{D^+}$ at the physical
light-quark mass point. \label{fig:chiral-fit}}
\end{figure}

\Figref{chiral-fit} shows our central fit to partially quenched data at all four lattice spacings.  The fit includes additional data 
(not shown) from ensembles at $a\approx0.12\;$fm ($\beta=6.0$) with $m_s$ lighter than physical
or with volumes $24^3\times 64$ and
$40^3\times64$ (see \tabref{ensembles}), which were generated to check finite volume effects.   We then extrapolate
the parameters to the continuum,  
adjust the  strange
sea-quark mass and charm valence- and sea-quark masses to their physical values, and set the light sea-quark mass equal to the light valence mass
(up to the small difference between $m_d$ and $m_l = (m_u+m_d)/2$)
giving the orange band.  Putting in the physical light quark mass then gives the black burst, which is 
the result for $\Phi_{D^+}$.  Note that the effect of isospin violation in the valence quarks is included in our
result; isospin violation in the sea is not included but is negligible.
 The biggest source of variation in the data in these four plots is not discretization errors,
but mistunings of the strange and, most importantly, charm quark mass.

The statistical error in $\Phi_{D^+}$ and $\Phi_{D_s}$ given by the fit in \figref{chiral-fit} is only part of the total statistical
error, since it does not include the statistical errors in the inputs of quark masses and lattice scale.  To determine the
total statistical error of each output quantity, we divide the full data set into 100 jackknife subensembles.  The complete calculation, including the
determination of the inputs, is performed on each subensemble, and the error is computed as usual from the variations over the subensembles.  Each subensemble drops approximately 10 consecutive stored configurations (50 to 60 trajectories) from each (completed) ensemble. This procedure  controls for autocorrelations, since all our measures of the autocorrelations of these quantities indicate that they are negligible after 4 or 8 consecutive configurations.  For the incomplete
physical-mass 0.06 fm ensemble with 583 configurations, we are forced to drop only about 6
consecutive stored configurations at a time.  Our expectation is that the effect of any remaining autocorrelations, while perhaps not completely negligible, is small compared to other sources of error.    
The total statistical errors computed from the jackknife procedure are only about 10\% larger than the statistical error from the chiral/continuum fit, indicating that the inputs are
are statistically quite well determined.

\begin{figure}[t]
\begin{center}
\null\vspace{-08mm}
\includegraphics[width=13cm]{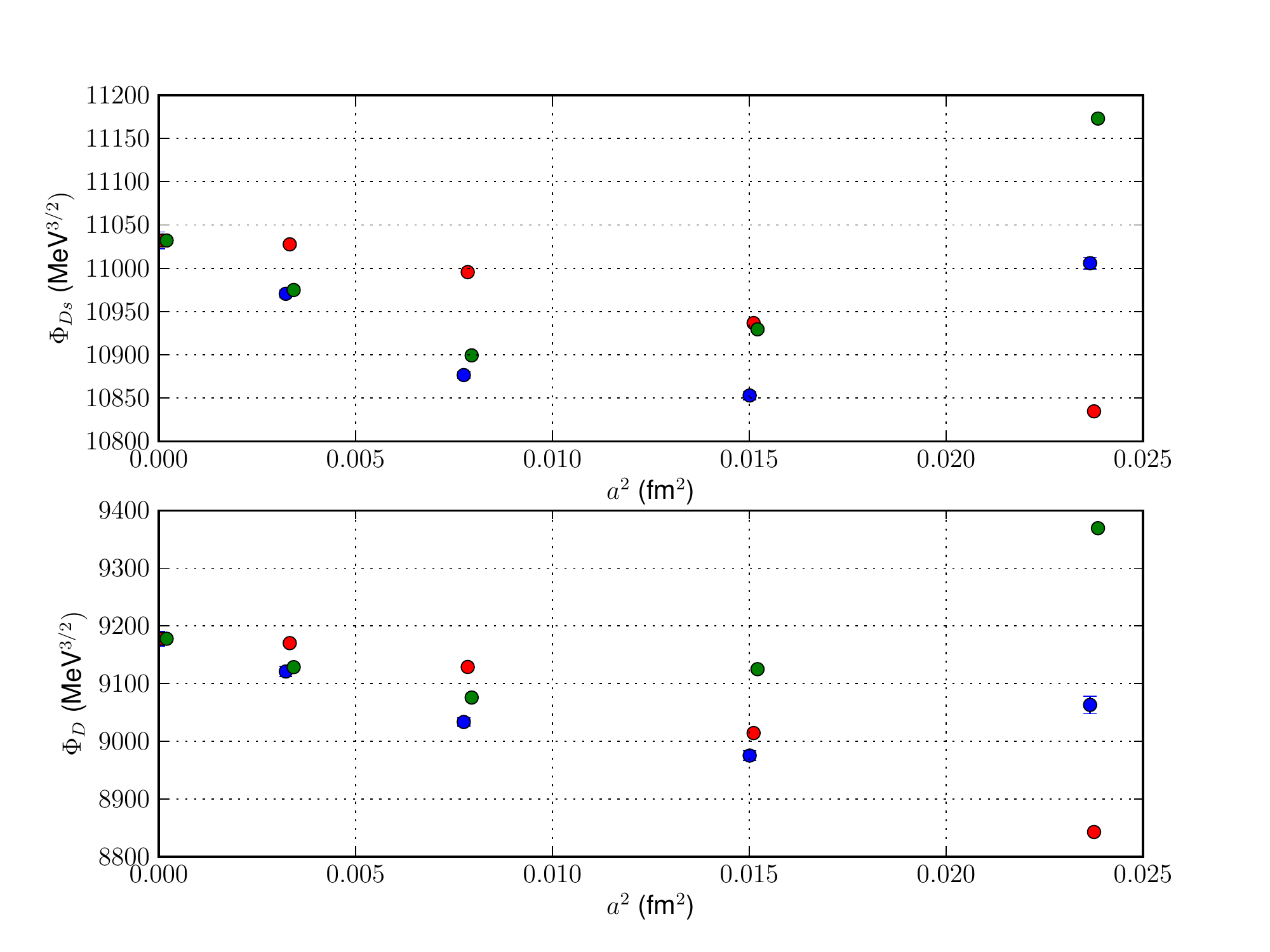}
\end{center}
\vspace{-8mm}
\caption{
Lattice spacing dependence of $\Phi_{D^+}$ and $\Phi_{D_s}$. The blue points show
the lattice data, after adjustment for mistunings of valence-and sea-quark masses.  The red points
show the contribution from the chiral logarithms, while the green points show the $a^2$ dependence induced by the fit parameters. 
Red and green points overlap at $a=0$ (only the green is visible).  \label{fig:a2dep}}
\end{figure}

\Figref{a2dep} illustrates how data for $\Phi_{D^+}$ and $\Phi_{D_s}$ depend on lattice spacing after 
adjustment to physical values of the quark masses (blue points).  There is a 2--3\% variation between 
these points and the continuum value (green point at $a^2=0$).  Note that there is clear curvature
in the plot, evidence of significant $a^4$ terms in addition to the formally leading $\alpha_s a^2$ terms.
The red points show the contribution from the chiral logarithms (with known taste splittings) to the $a^2$ dependence 
of the chiral fit function. The green points show the corresponding contribution from the analytic fit parameters.
The two effects are of comparable magnitudes but the relative sign changes with lattice spacing; both are needed to describe 
the  $a^2$ dependence of the data.    

\section{Continuum extrapolation and systematic errors}\vspace{-2.0mm} \label{SYSERRS}

\begin{figure}
\vspace{-0.70in}
        \begin{center}\begin{tabular}{ll}
\hspace{-0.3in} \includegraphics[width=0.55\textwidth]{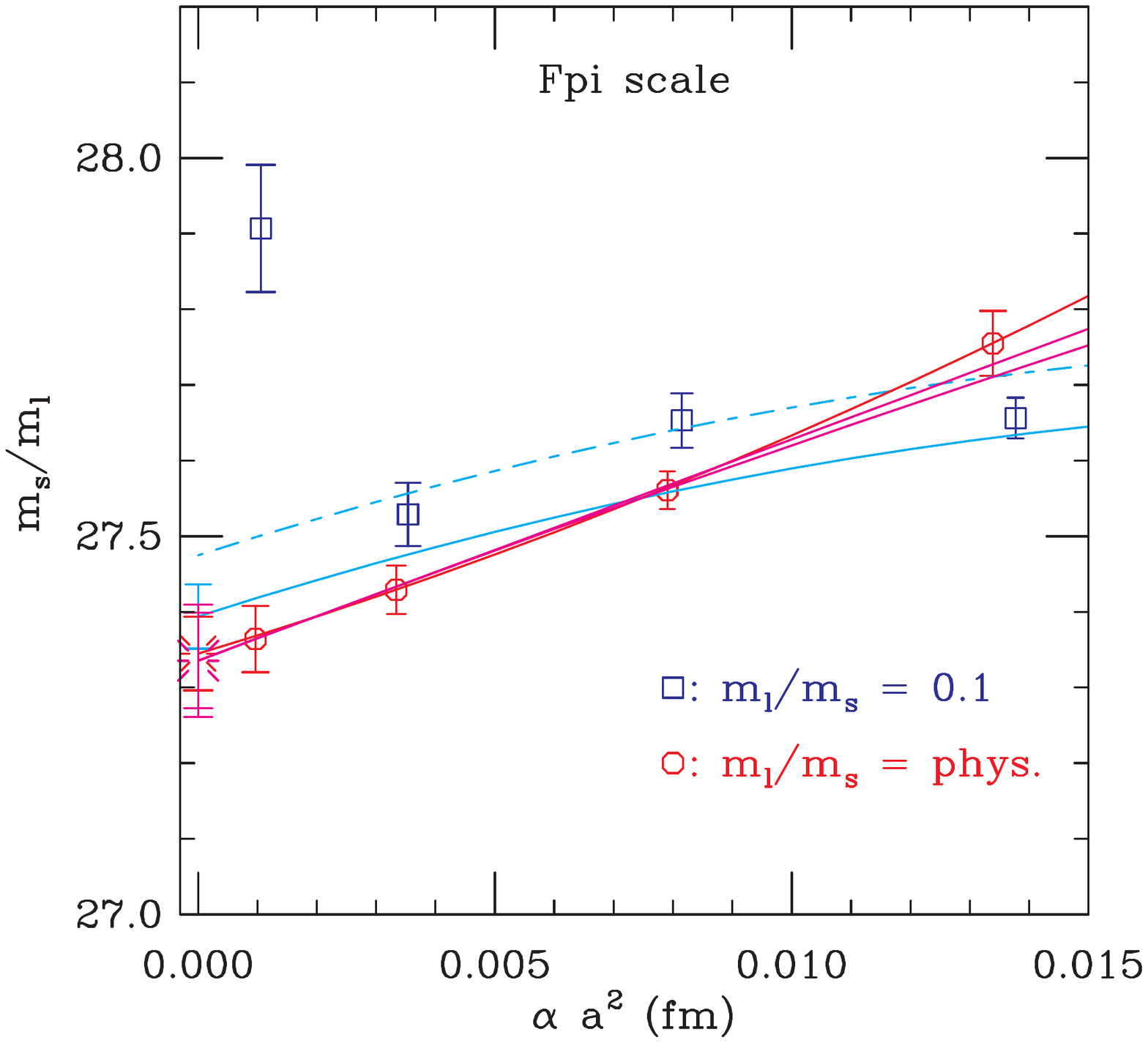} &
\hspace{-0.4in} \includegraphics[width=0.55\textwidth]{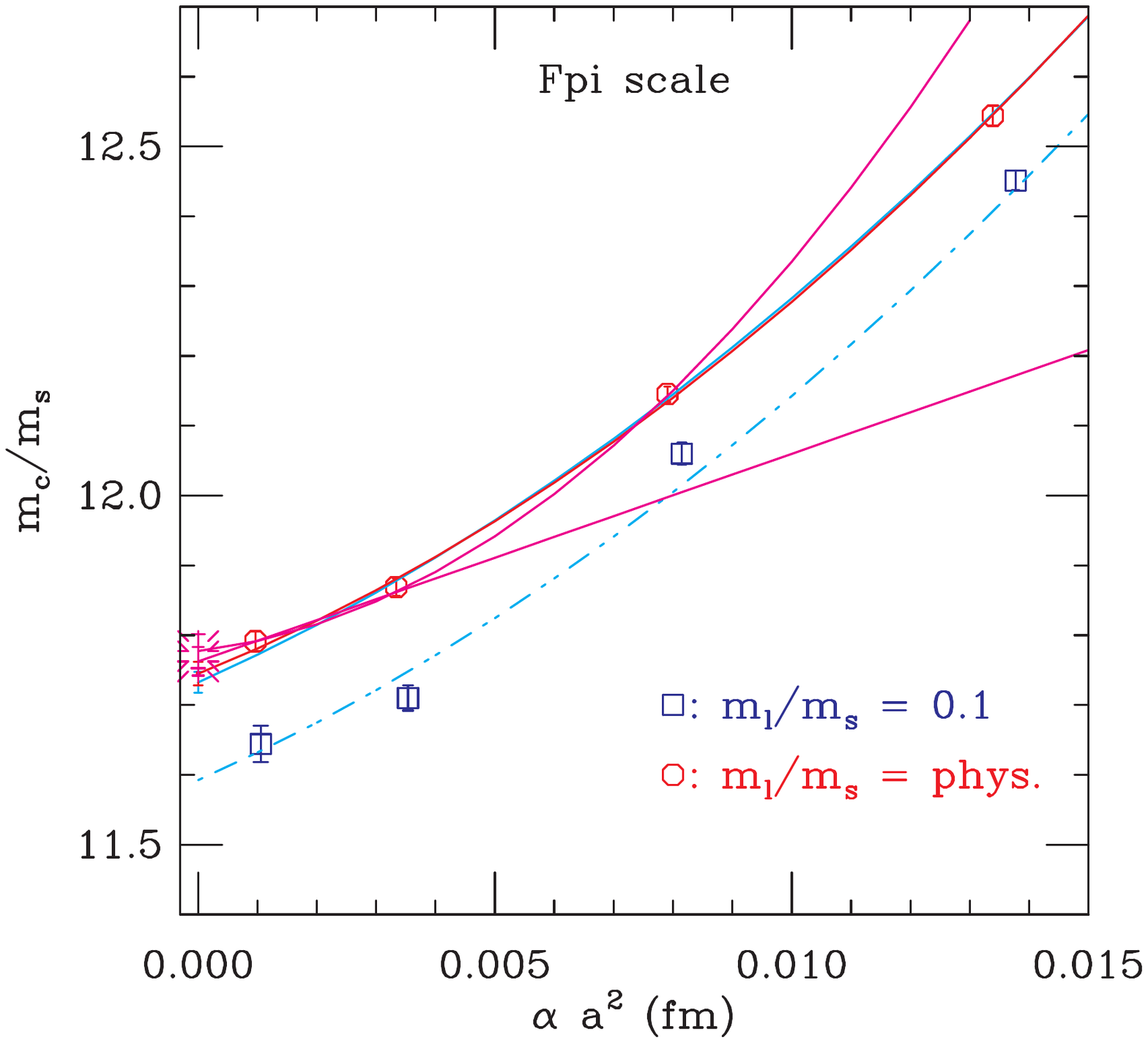} \\
        \end{tabular}\end{center}
\vspace{-1.1in}
\caption{\label{fig:quarkmass_ratios}
Quark mass ratios $m_s/m_l$ and $m_c/m_s$ on the physical and $0.1\,m_s$ ensembles,
together with fits and continuum extrapolations described in the text.  The red line is our preferred fit,
which is quadratic in $a^2 \alpha_{TV}$, determined from taste violations in the pion
masses, using the physical quark mass ensembles with small corrections for valence
quark mass mis-tuning.  The cyan lines are a fit quadratic in $a^2 \alpha_{TV}$ and linear
in light valence mass, including the $0.1\,m_s$ ensembles.  There the solid line is the
fit evaluated at the physical light quark mass and the dashed line is the fit evaluated
at $0.1\,m_s$.  The magenta lines are zero degree-of-freedom fits to the physical ensembles,
using a quadratic through the lowest three points or a line through the lowest two.  Other
fits not shown here are quadratic in $a^2 \alpha_V$ or $a^2$.
}
\vspace{-0.05in}
\end{figure}

Our treatment of finite volume effects in the light quark meson masses and decay
constants is described in Ref.~\cite{LATTICE12_FD}.  To summarize, using NLO staggered chiral
perturbation theory and results from three ensembles differing only in spatial size,
we found the values that $f_\pi$, $M_\pi$, $f_K$ and $M_K$ would take in a 5.5 fermi box, the
size of our physical quark mass ensembles.  These adjusted values were then used in the
tuning procedure described above.  Afterwards, when $f_K$ is reported, we reverse this
correction to get the value at infinite volume. We use the difference between NLO staggered
chiral perturbation theory and NNLO continuum chiral perturbation theory as an estimate of
the remaining finite volume errors in the tuning procedure.

Electromagnetic effects in the kaon masses used in the tuning procedure were treated
by using kaon masses adjusted for EM effects in the tuning procedure \cite{EM_EFFECTS}.
Remaining errors from these EM effects were estimated by varying the parameter that
characterizes violation of Dashen's theorem, $\Delta_{EM}$, by one standard deviation and
by varying the less well understood shift in the average squared kaon mass by one
half of its estimated value of $900$ MeV$^2$. (The latter estimate does not include the effects of
EM quark mass renormalization, which is not determined in the calculation of Ref.~\cite{EM_EFFECTS}.)

For most quantities, our largest systematic error is the error from the continuum
extrapolation and quark mass adjustments.  In the tuning procedure which finds the
valence-quark masses, and in the determination of $f_K/f_\pi$, this was done by trying 
several different fits, as illustrated in Fig.~\ref{fig:quarkmass_ratios} for $m_s/m_l$ and $m_c/m_s$.
Our first fit was quadratic in $\alpha_{TV} a^2$, where $\alpha_{TV}$ is proportional to
a coupling constant determined from mass splittings among the different pion tastes,
 and linear in sea-quark $m_l/m_s$, using the physical quark mass
and $m_l/m_s=1/10$ ensembles.
Central values for these quantities come from a fit that
is quadratic in $\alpha_{TV} a^2$, using the physical quark mass ensembles with small
adjustments  (using the coefficient of $m_l/m_s$ from the first fit) for sea-quark mass mistuning.
We also considered fits to the physical mass ensembles that were quadratic in $\alpha_V a^2$ or
$a^2$, where $\alpha_V$ is determined from the plaquette.  Finally, we considered
quadratic and linear extrapolations of the physical quark mass ensembles using
 the finest three and two lattice spacings respectively.  We used the full range
of variation among these extrapolations as the continuum extrapolation systematic
error on the quark mass ratios and on $f_K/f_\pi$.

\begin{figure}[t]
\begin{center}
\null\vspace{-48mm}
\begin{tabular}{l l}
\null\hspace{-20mm}\includegraphics[width=10.7cm]{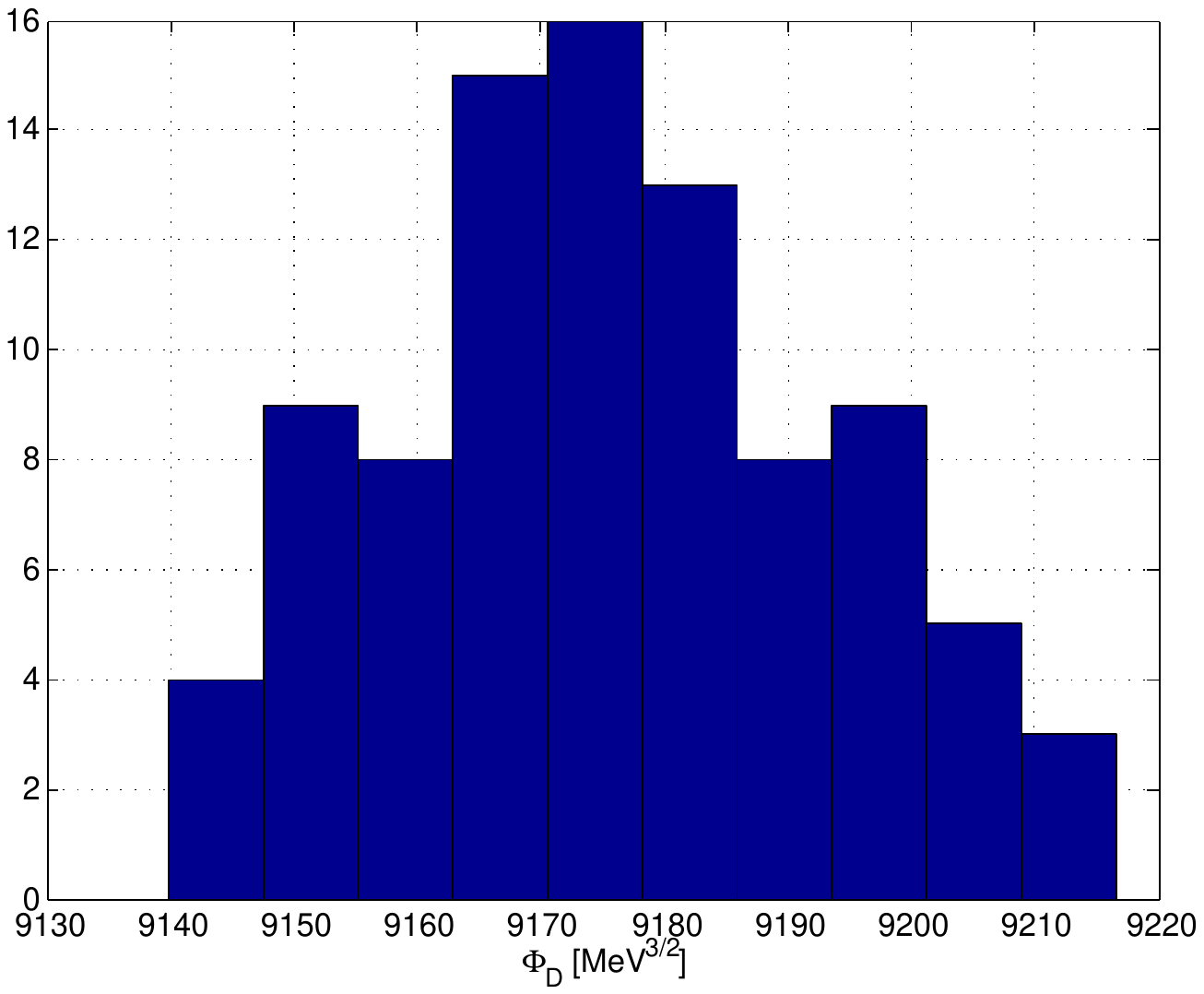}
&\hspace{-37mm}
\includegraphics[width=10.7cm]{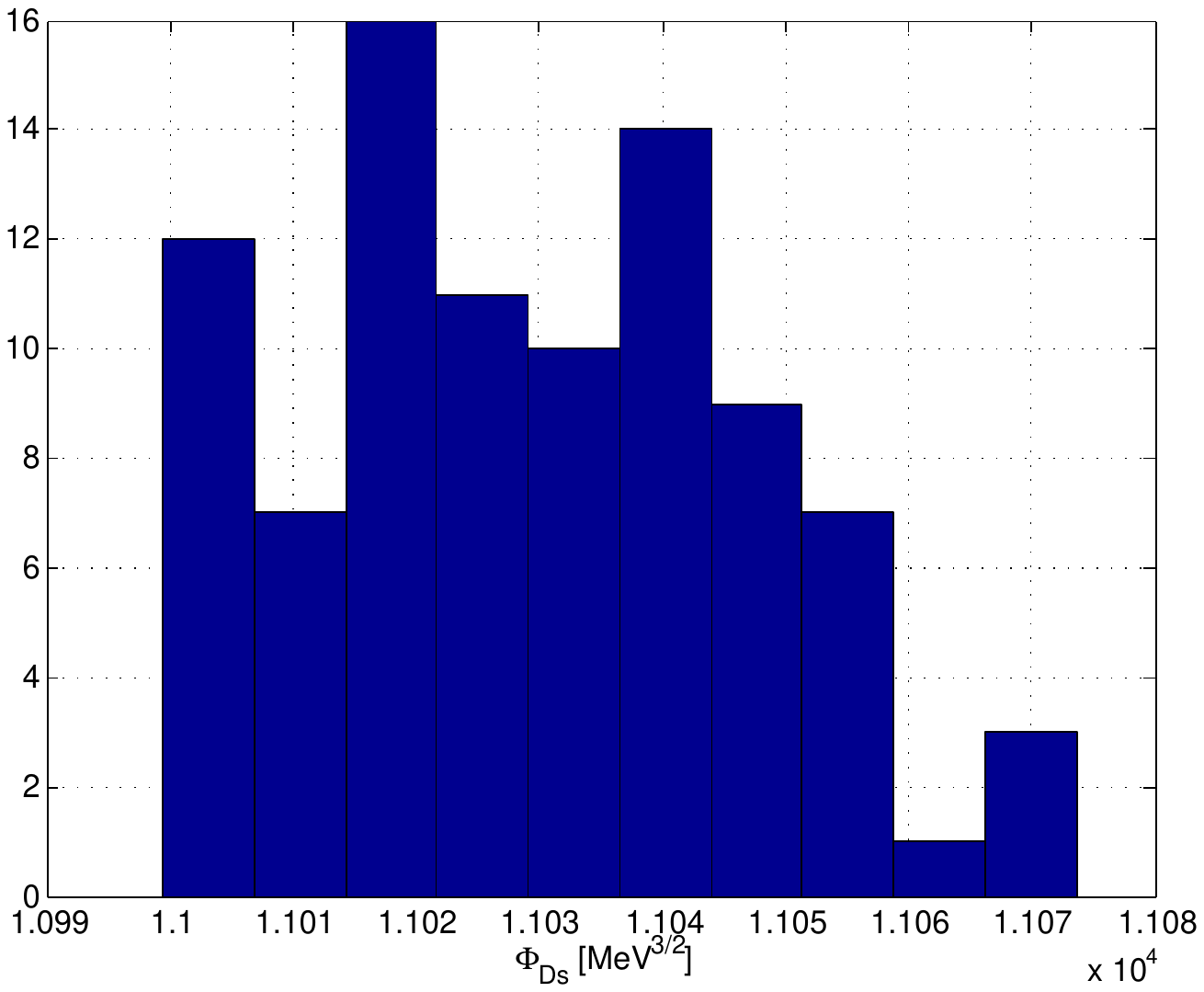}
\end{tabular}
\end{center}
\vspace{-43mm}
\caption{
Histograms of $\Phi_{D^+}$ and $\Phi_{D_s}$ values obtained from various versions of the continuum/chiral extrapolation and various inputs of quark masses and scale values from the physical-mass analysis. Our central fit gives $\Phi_{D^+}=9178\; {\rm MeV}^{3/2}$ and $\Phi_{D_s}=11032\; {\rm MeV}^{3/2}$. \label{fig:hist_phi}}
\end{figure}

To determine the systematic error associated with the continuum extrapolation (and chiral interpolation) 
of the charmed decay constants in the chiral perturbation theory analysis, we rerun the analysis
with alternative continuum/chiral fits, and with alternative inputs that come from different continuum extrapolations of the
physical-mass analysis. 
We have a total of
15 acceptable ($p\ge0.05$) versions of the continuum/chiral fits, which keep or drop the $a\approx0.15$ fm ensemble,
keep or drop $\alpha_s(am_c)^2$ and $(am_c)^4$ terms, constrain higher order chiral terms and/or discretization terms
with priors or leave them unconstrained, add or omit additional parameters that permit the taste-violating parameters
$L_a$, $\delta'_A$ and $\delta'_V$ to vary differently with lattice spacing than simply as $\bar \Delta$, \etc We also have the six versions
of the continuum extrapolations used in the tuning procedure that leads to the inputs of quark mass and lattice scale.
This gives a total of 90 versions
of the analysis.  Histograms of the 90 results for $\Phi_{D^+}$ and $\Phi_{D_s}$  are shown in \figref{hist_phi}.  Conservatively, 
we take the maximum difference seen in these results with our central values as the ``self-contained'' estimate of the continuum 
extrapolation errors within this chiral analysis. We now have carried out additional fits, beyond those shown at the conference, resulting in small increases in our estimated errors on some of the quantities. We now 
also  choose a central fit that gives 
results close to the centers of the histograms, which results in more symmetrical error bars.

In practice, the NLO finite volume corrections are included in our fit function, \eq{chiral-form}, when it is applied to the data,
and the volume is sent to infinity when the continuum results are extracted.  We may conservatively
estimate the residual finite volume error
in the heavy-light data either by turning off all finite volume corrections and repeating the fit, or by using the current fit to
find the size of the NLO finite volume correction on our most-important, 0.06 fm physical-mass ensemble. Yet another way to
make the estimate is by direct comparison of our results on the $32^3\times 64$, $\beta=6.0$,
$m_l=0.1m_s$ ensemble
(which is similar in physical size to our other $m_l=0.1m_s$ ensembles) and the $40^3\times 64$, $\beta=6.0$,
$m_l=0.1m_s$ ensemble. All three methods indicate that there are negligible direct finite volume effects in the heavy-light lattice data. Nevertheless, there are non-negligible finite volume effects in our final answers, which appear due to the scale setting in the light-quark sector through, ultimately, $f_\pi$.  (The value of $f_{p4s}$ in physical units that we use comes by comparison with
$f_\pi$.)  We then propagate the errors in  the inputs through our analysis.  Electromagnetic errors in the light quark masses are similarly propagated through our analysis.

\section{Results}\vspace{-2.0mm} \label{RESULTS}

The results for $D$-meson  decay constants from the self-contained chiral fit analysis are:
\begin{eqnarray}
f_{D^+} &=& 212.3 \pm0.3_{\rm stat}\;\pm0.9_{a^2\,{\rm extrap}}\pm 0.3_{\rm FV}\pm 0.0_{\rm EM}\pm 0.3_{f_\pi\, {\rm expt}}    \ {\rm MeV}\eqn{fD-chiral-result}\\
f_{D_s} &=& 248.7\pm 0.2_{\rm stat}\;{}^{+0.9}_{-0.7}\vert_{a^2\,{\rm extrap}}\pm 0.2_{\rm FV}\pm0.1_{\rm EM} \pm0.4_{f_\pi\, {\rm expt}}  \  {\rm MeV} \eqn{fDs-chiral-result}\\
f_{D_s}/f_{D^+} &=& 1.1714(10)_{\rm stat}({}^{+23}_{-21})_{a^2\,{\rm extrap}}(3)_{\rm FV}(5)_{\rm EM}
\eqn{ratio-chiral-result}
\end{eqnarray} 
It should be emphasized that the EM errors we estimate here, which come from effects on light-meson, and hence light-quark, masses, do not include the EM effects on the $D_s$ mass, which we use to fix the $c$-quark mass.  
 We make a rough estimate of this omitted EM effect below.  
 In addition, we note that we are computing the values the decay constants as they are conventionally defined,
in a pure-QCD world. Comparison to experiment thus requires a matching of the decay rates between
 QCD and QCD+QED.  The errors in such a matching are not included in our error estimates.

We take the central values and statistical errors in \eqsthru{fD-chiral-result}{ratio-chiral-result} as our best
estimates.  For the continuum extrapolation error,
we also consider the differences of the central values with corresponding quantities obtained by various
continuum extrapolations of the  straightforward analysis on the
physical-mass ensembles,  and take those differences as the error whenever they are larger
than the self-contained error.   The analysis on the physical-mass ensembles also gives alternative error
estimates for the finite volume and EM errors, which however turn out to be the same as, or slightly smaller than, those in   \eqsthru{fD-chiral-result}{ratio-chiral-result}.
This procedure gives our current best (but still preliminary) 
results for $f_{D^+}$, $f_{D_s}$ and $f_{D_s}/f_{D^+}$.
We also include results for quark-mass ratios coming from the tuning procedure described 
in Sec.~\ref{sec:by_ensemble}, and $f_{K^+}/f_\pi$ coming from the ensemble-by-ensemble analysis:
\begin{eqnarray}
m_c/m_s &=& 11.741(19)_{\rm stat}(59)_{\rm sys} \\
m_s/m_l &=& 27.366(52)_{\rm stat}(107)_{\rm sys} \\
m_u/m_d &=& 0.4619(48)_{\rm stat}(169)_{\rm sys} \\
f_{K^+}/f_\pi &=& 1.1957(10)_{\rm stat}\pm(25)_{\rm sys} \\
f_{D^+} &=& 212.3 \pm 0.3_{\rm stat}\;{}^{+0.9}_{-1.1}\vert_{a^2\,{\rm extrap}}\pm 0.3_{\rm FV}\pm 0.0_{\rm EM}\pm 0.3_{f_\pi\, {\rm expt}}    \ {\rm MeV}\eqn{fD-result}\\
f_{D_s} &=& 248.7\pm  0.2_{\rm stat}\;{}^{+0.9}_{-1.4}\vert_{a^2\,{\rm extrap}}\pm 0.2_{\rm FV}\pm0.1_{\rm EM} \pm0.4_{f_\pi\, {\rm expt}}  \  {\rm MeV} \\
f_{D_s}/f_{D^+} &=& 1.1714(10)_{\rm stat}({}^{+29}_{-21})_{a^2\,{\rm extrap}}(3)_{\rm FV}(5)_{\rm EM}
\eqn{ratio-result}
\end{eqnarray}
The EM errors here do not include the effect of the EM contribution to  the $D_s$ mass, which has not been directly determined
in QCD+QED simulations.  This effect would change  $m_c/m_s$ and, in turn, $f_{D^+}$ and $f_{D_s}$.  We may roughly estimate the effect by assuming the EM effect on $M_{D_s}$ is similar to that on $m_K$, namely a few MeV. 
This would change 
$m_c/m_s$ by $\sim\!0.01$--0.03,  $f_{D^+}$ and $f_{D_s}$  by $\sim\!0.01$--0.05 MeV, and
$f_{D_s}/f_{D^+}$  $\sim\!0.0001$--0.0003.

\begin{figure}
\vspace{-0.0in}
        \begin{center}\begin{tabular}{ll}
\hspace{-0.25in} \includegraphics[width=0.50\textwidth]{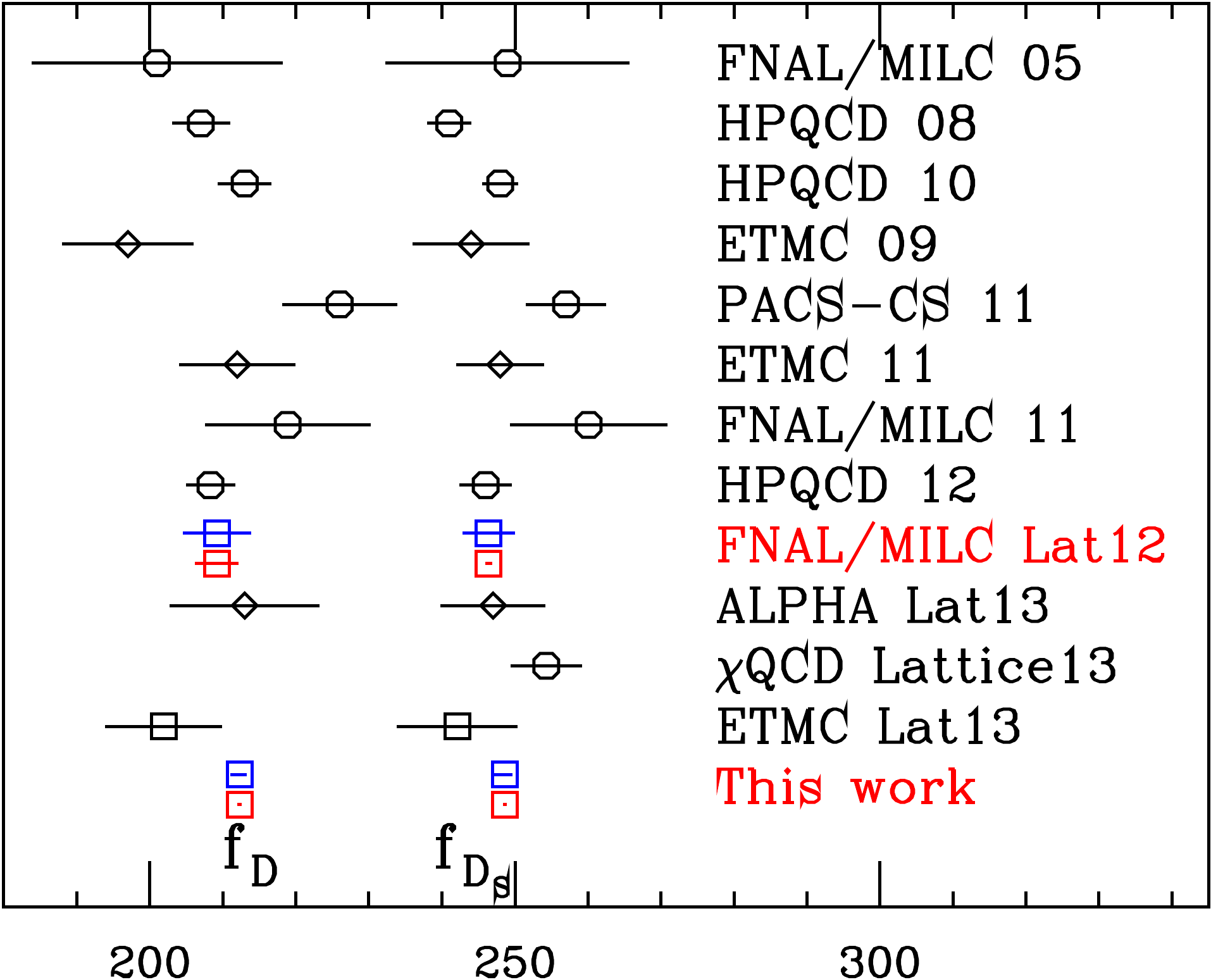} &
\hspace{0.2in} \includegraphics[width=0.43\textwidth]{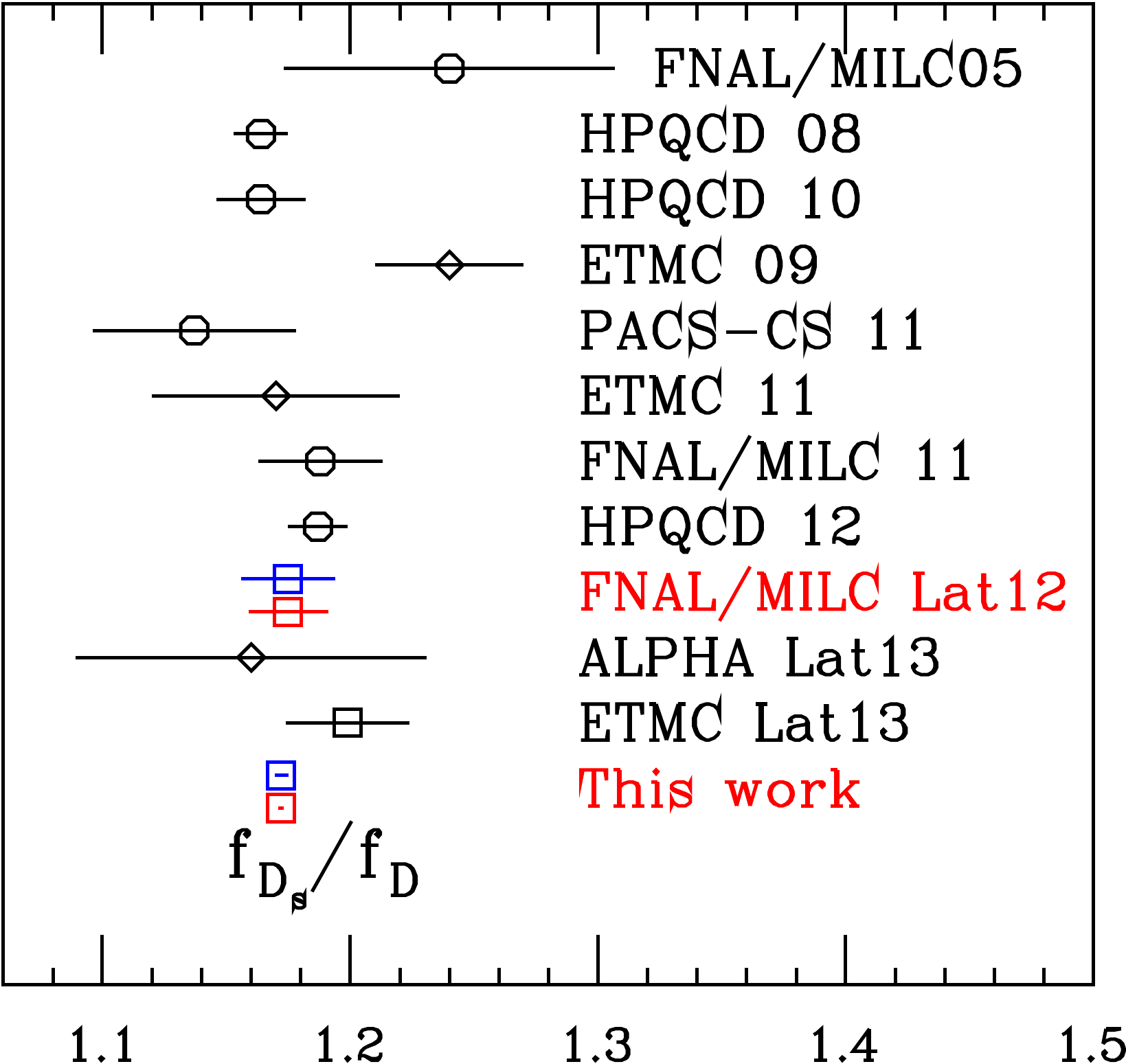} \\
        \end{tabular}\end{center}
\vspace{-0.22in}
\caption{\label{fig:fdds_values}
Lattice computations of $f_D$, $f_{D_s}$ and $f_{D_s}/f_D$.
Diamonds are $n_f=2$ calculations, octagons $n_f=2+1$ and squares $n_f=2+1+1$.
Earlier work is from references
\protect\cite{FNAL95,HPQCD08,HPQCD10,ETMC09,PACS-CS11,ETMC11,FNAL11,HPQCD12,
CHIQCD13,ALPHA13,ETMC13}.
In the results from these proceedings and the FNAL/MILC Lattice 12 results,
\R{red} points have statistical errors only, \B{blue} include systematic errors.
}
\end{figure}

\begin{figure}
\vspace{-1.40in}
\begin{center}\includegraphics[width=0.72\textwidth]{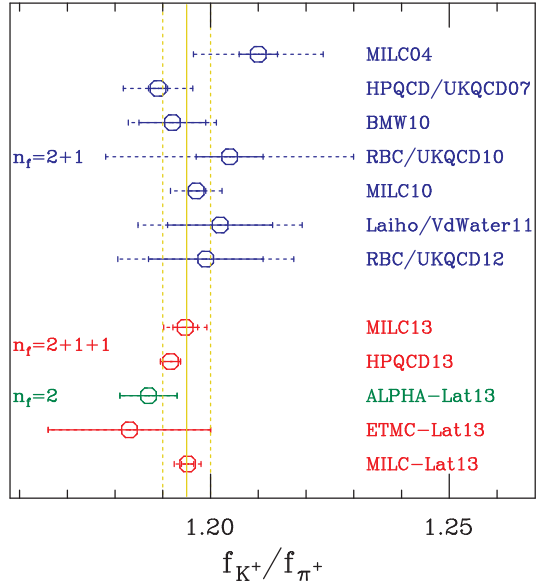}\end{center}
\vspace{-1.25in}
\caption{ \label{fig:fkofpivalues}
Determinations of $f_K/f_\pi$.
Earlier work is from
references~\cite{FKOFPI_MILC13,ETMC13,FKOFPI_MILC04,FKOFPI_HPQCD-UKQCD07,FKOFPI_MILC10,FKOFPI_RBC-UKQCD10,FKOFPI_BMW10,FKOFPI_HPQCD13,FKOFPI_ALPHA13}.
The vertical lines show the central value and errors of the FLAG 2+1+1 average \cite{Aoki:2013ldr}. 
}
\end{figure}

These decay constants have been computed by several groups using a variety of gauge
and fermion actions.  In Figs.~\ref{fig:fdds_values} and \ref{fig:fkofpivalues} we compare the
results in this work with previous computations of the decay constants.

\acknowledgments
This work was supported by the U.S. Department of Energy and National
Science Foundation,
by the URA Visiting Scholars' program (A.E-K.),
and by the MINECO, Junta de Andaluc\'{\i}a, and European Commission.
Computation for this work was done at
the Argonne Leadership Computing Facility (ALCF),
the National Center for Atmospheric Research (UCAR), 
Bluewaters at the National Center for Supercomputing Resources (NCSA),
the National Energy Resources Supercomputing Center (NERSC),
the National Institute for Computational Sciences (NICS),
the Texas Advanced Computing Center (TACC),
and the USQCD facilities at Fermilab,
under grants from the NSF and DOE.


\begin{thebibliography}{99}

\bibitem{HPQCD_HISQ}
E.~Follana {\it et al.} [HPQCD Collaboration],
Phys.\ Rev.\ D {\bf 75}, (2007) 054502 [hep-lat/0610092].

\bibitem{HISQ_CONFIGS}
A.~Bazavov {\it et al.} [MILC Collaboration],
Phys.\ Rev.\ {\bf D87}, 054505 (2013),
[arXiv:1212.4768].

\bibitem{KIM_LATTICE12}
A.~Bazavov {\it et al.},
PoS(Lattice 2012)158,
[arXiv:1212.0613].

\bibitem{EM_EFFECTS}
L.~Levkova {\it et al.} [MILC Collaboration], \pos{PoS(Lattice2012)137};
C.~Bernard {\it et al.} [MILC Collaboration], \pos{PoS(CD12)030} 
[arXiv:1301.7137].

\bibitem{Bernard-Komijani}
J.\ Komijani and C.\ Bernard, 
PoS LAT2012 (2012) 199 [arXiv:1211.0785];
C.\ Bernard and J.\ Komijani, 
[arXiv:1309.4533].


\bibitem{FermilabMILC_Dec2011}
  A.~Bazavov
  {\it et al.} [Fermilab Lattice and MILC Collaborations],
Phys.\ Rev.\ D {\bf 85}, 114506 (2012) [arXiv:1112.3051].

\bibitem{BoydGrinstein}
C. G. Boyd and B. Grinstein, Nucl.\ Phys.\ {\bf B442}, 205 (1995),
[arXiv:hep-ph/9402340].

\bibitem{Aubin:StagHL2007}
C.\ Aubin and C.~Bernard,
Phys.\ Rev.\ D76, 014002 (2007),
[arXiv:hep-lat/0704.0795].


  \bibitem{Bazavov:2011fh}
  A.~Bazavov {\it et al.}  [MILC Collaboration],
  PoS LATTICE {\bf 2011}, 107 (2011)
  [arXiv:1111.4314 [hep-lat]], and work in progress.

\bibitem{FKOFPI_MILC13}
A.~Bazavov {\it et al.} [MILC Collaboration],
Phys.\ Rev.\ Lett.\ {\bf 110}, 172003 (2013).

\bibitem{Becirevic}
D.\ Becirevic and F.\ Sanfilippo, Phys.\ Lett.\ {\bf B721}, 94 (2013) [arXiv:1210.5410];
K.\ U.\ Can, G.\ Erkol, M.\ Oka, A.\ Ozpineci, and T.\ T.\ Takahashi, Phys.\ Lett.\ {\bf B719}, 103
 (2013) [arXiv:1210.0869].
 
 \bibitem{Detmold}
 W.\ Detmold, C.\ J.\ D.\ Lin, and S.\ Meinel, Phys.\ Rev.\ D{\bf 85}, 114508 (2012) [arXiv:1203.3378].

\bibitem{LATTICE12_FD}
A.~Bazavov {\it et al.} [Fermilab Lattice and MILC Collaborations],
PoS(Lattice 2012)159 [arXiv:1210.8431].



\bibitem{FNAL95}
C.~Aubin {\it et al.} [Fermilab Lattice and MILC Collaborations], Phys.\ Rev.\ Lett.\ {\bf 95}, 122002(2005),
[arXiv:hep-lat/0506030].

\bibitem{HPQCD08}
E.~Follana {\it et al.} [HPQCD Collaboration], Phys.\ Rev.\ Lett.\ {\bf 100}, 062002 (2008) [arXiv:0706.1726].

\bibitem{HPQCD10}
C.~Davies {\it et al.} [HPQCD Collaboration], Phys.\ Rev.\ D {\bf 82}, 114504 (2010) [arXiv:1008.4018].

\bibitem{ETMC09}
B.~Blossier {\it et al.} [ETM Collaboration], JHEP {\bf 0907} 043 (2009) [arXiv:0904.0954].

\bibitem{PACS-CS11}
Y.~Namekawa {\it et al.} [PACS-CS Collaboration], Phys.\ Rev.\ D {\bf 84} (2011) 074505 [arXiv:1104.4600].

\bibitem{ETMC11}
P.~Dimopoulos {\it et al.} [ETM Collaboration], JHEP {\bf 01}, 046 (2012) [arXiv:1107.1441].

\bibitem{FNAL11}
J.A.~Bailey {\it et al.} [Fermilab Lattice and MILC Collaborations], PoS(Lattice2011)320 [arXiv:1112.3978];
A.~Bazavov {\it et al.} [Fermilab Lattice and MILC Collaborations], Phys.\ Rev.\ D {\bf 85} 114506 (2012),
[arXiv:1112.3051].

\bibitem{HPQCD12}
H.~Na {\it et al.} [HPQCD Collaboration], Phys.\ Rev.\ D {\bf 85} 125029 (2012) [arXiv:1206.4936].

\bibitem{CHIQCD13}
Y.~Yang, Y.~Chien, Z.~Liu, these proceedings

\bibitem{ALPHA13}
J.~Heitger, G.~von Hippel, S.~Schaefer and F.~Virotta [ALPHA Collaboration],
these proceedings.

\bibitem{ETMC13}
L.~Riggio [ETM Collaboration], these proceedings.


\bibitem{FKOFPI_MILC04}
C.~Aubin {\it et al.} [MILC Collaboration]
Phys.\ Rev.\ {\bf D} 70, 114501 (2004). 

\bibitem{FKOFPI_HPQCD-UKQCD07}
E.\ Follana, C.\ T.\ H.\ Davies, G.\ P.\ Lepage and J.\ Shigemitsu [HPQCD/UKQCD Collaboration],
Phys.\ Rev.\ Lett.\ {\bf 100} (2008) 062002  [arXiv:0706.1726].

\bibitem{FKOFPI_MILC10}
A.~Bazavov {\it et al.} [MILC Collaboration],
PoS LATTICE2010, 074 (2010). 

\bibitem{FKOFPI_RBC-UKQCD10}
Y.\ Aoki \et\ [RBC/UKQCD Collaboration],
Phys.\ Rev.\ D{\bf 83} (2011) 074508 [arXiv:1011.0892].

\bibitem{FKOFPI_BMW10}
S.~D\"urr \et\ [BMW Collaboration],
Phys.\ Rev.\ D {\bf 81}, 054507 (2010)
[arXiv:1001.4692].


\bibitem{FKOFPI_HPQCD13}
R.J.~Dowdall, C.T.H.~Davies, G.P.~Lepage and C.~McNeile [HPQCD Collaboration],
[arXiv:1303.1670]; R. Dowdall {\it et al.}, these proceedings.

\bibitem{FKOFPI_ALPHA13}
S.~Lottini [ALPHA Collaboration], these proceedings

\bibitem{Aoki:2013ldr} 
  S.~Aoki  {\it et al.} [FLAG],
  arXiv:1310.8555 [hep-lat].
\end{thebibliography}
\end{document}